%% file: nloee.tex
\documentclass{JHEP3}
\usepackage{amsmath}
\usepackage{axodraw}
\usepackage{epsfig}
\usepackage{amssymb}
\usepackage{mathrsfs}
\usepackage{mathcomp}
\usepackage{cite}
\usepackage{stmaryrd}
\usepackage[latin1]{inputenc}

\input mydefs

\def\ie{\emph{i.e.}}
\def\eg{\emph{e.g.}}

\def\yms{\ensuremath{y_{\mbox{\tiny MS}}}}
\def\ycut{\ensuremath{y_{\mrm{cut}}}}

\def\lqcd{\Lambda_{\mrm{QCD}}}

\newcommand{\alpgen}{A\scalebox{0.8}{LPGEN}\xspace}

\preprint{LU-TP 08-20\\MCnet/08/13}

\title{Extending CKKW-merging to One-Loop Matrix Elements\footnote{Work
    supported in part by the Marie Curie RTN ``MCnet'' (contract
    number MRTN-CT-2006-035606).}}

\author{Nils Lavesson and Leif Lönnblad\\
  Dept.~of Theoretical Physics,
  Sölvegatan 14A, S-223 62  Lund, Sweden\\
  E-mail: \email{Nils.Lavesson@thep.lu.se}
    and \email{Leif.Lonnblad@thep.lu.se}}

  \abstract{ We extend earlier schemes for merging tree-level matrix
    elements with parton showers to include also merging with one-loop
    matrix elements. In this paper we make a first study on how to
    include one-loop corrections, not only for events with a given jet
    multiplicity, but simultaneously for several different jet
    multiplicities. Results are presented for the simplest non-trivial
    case of hadronic events at LEP as a proof-of-concept.  }

\begin{document}

\sloppy

\section{Introduction}

One of the big theoretical challanges with LHC physics is
the description of states with many hard jets. The high energy and
large rapidity range for jets at the LHC means that producing
multi-jet events from QCD processes is more likely than ever
before. These states compose the background for many of the channels
that could contain new physics. It is therefore important to get as
good description as possible for the multi-jet states within QCD.

Monte Carlo event generators have become standard tools for
simulating events in a particle collider. These event generators try
to simulate events with all the characteristics of a real event. The
theoretical basis is a parton shower, which is combinded with
hadronization models to produce the final-state hadrons. The
hadronization models are phenomenological models that only work
reliably when all the partons in the soft and collinear limit have
been simulated corretly in the shower. Parton showers are based on
expanding the emission probabilities in this limit, which makes them
suitable to use together with the hadronization models.  Although
parton showers have been used to describe a wide range of results with
good accuracy, it is well known that they cannot give a good
description of observables sensitive to emissions away from the
collinear and soft regions.

The way to improve the description of multi-jet states is to include
exact matrix elements. The matrix elements describe these states well,
but do not provide a way of describing emissions in the soft or
collinear limit. In fact, the matrix elements are divergent in these
limits and have to be regulated using a cutoff. To correctly describe
the final-state hadrons in multi-jet events, the matrix elements
and the parton shower descriptions need to be combined. This has been
done for tree-level matrix elements using algorithms such as
CKKW\cite{Catani:2001cc, Krauss:2002up}, CKKW-L\cite{Lonnblad:2001iq,
  Lavesson:2005xu}, MLM \cite{MLM, Mangano:2006rw} and Pseudo-Shower
\cite{Mrenna:2003if}.

In recent years a lot of effort has been put into calculating one-loop
matrix elements to be able to predict observables to next-to-leading
order (NLO) accuracy. There are several program available to do this,
\eg\ MCFM\cite{mcfm} and NLOJET++\cite{Nagy:2003tz}. Currently efforts
are being made to automate the whole procedure and make more processes
available, including significantly higher parton muliplicities (MCFM
and BlackHat\cite{Berger:2008sj}). The one-loop matrix element
calculations contain important QCD contributions which cannot be
simulated with tree-level matrix elements nor with parton
showers. Preferably both the tree-level and one-loop matrix elements
as well as the parton shower should be used consistently
together. Lacking such a complete description, the uncertainties in a
given NLO calculation due to parton showers and hadronization are
typically estimated using a separate approximate Monte Carlo
simulation. Alternatively, for certain observables, it is possible to
combine a NLO calculation with analytically resummed parton-shower
correction together with a semi-universal power correction giving the
hadronization correction. However, it would clearly be advantegous if
the one-loop matrix elements could be used together with parton
showers and hadronization models in a more consistent manner.

A few algorithms have been presented to merge one-loop matrix elements
with parton showers. The two main algorithms are MC@NLO
\cite{Frixione:2002ik, Frixione:2006gn} and POWHEG \cite{Nason:2004rx,
  Frixione:2007vw}, but they are limited to only including the
one-loop matrix element for the lowest order process.  A similar
algorihtm for $e^+e^-$ to three jets was presented in
\cite{Kramer:2003jk, Soper:2003ya, Kramer:2005hw}. Other groups have
made proposals on how to go beyond this and include one-loop matrix
elements also for higher multiplicities\cite{Nagy:2007ty,
  Giele:2007di, Bauer:2008qh, Bauer:2008qj}\footnote{In particular
  \cite{Nagy:2007ty} is very close in spirit to the strategy presented
  here.}, but none have so far presented a complete implementation.

In this paper we present a general algorithm to include one-loop
matrix elements of any order in the strong coupling together with
parton showers. The idea is to take the first two terms in order of
$\as$ from the one-loop matrix element and all higher order terms from
the parton shower. Two different events samples are generated. The
first sample consists of events which are generated according to
one-loop matrix elements and dressed using a parton shower. The second
sample are events generated with a parton shower corrected with
tree-level matrix elements, where the first two terms in the
$\as$-expansion has been subtracted. The procedure is applied to all
the different multiplicities one wishes to calculate and in the end
all the samples are added.

Calculating the first two terms in orders of $\as$ in the shower
introduces complications because they include the first term in an
expansion of the Sudakov form factor. In addition the running \as\ used
in the shower also gives a contribution to the terms at this order.
These complications need to be dealt with in order to have a consistent
algorithm.

Our method uses the same philosophy as CKKW-L, which means that phase
space is split up in two different regions using a merging scale, and
the corrections to the matrix elements are simulated using the shower.
However, it should be noted that most of what is presented here could
also be used together with the CKKW algorithm, where the corrections are
calculated analytically. Using the shower rather than doing analytical
calculation means that non-leading terms, such as energy--momentum
conservation and recoil treatments, included in the shower, are also
included in the corrections to the matrix elements.

Although it would be more interesting to simulate jets at the LHC, we
limit ourselves to LEP physics in this paper. The reason is that the
inclusion of parton densities causes a number of additional
complications that needs to be studied further before an algorithm
valid also for incoming hadrons can be presented.

The outline of this paper is the following. In section
\ref{sec:theory} some key concepts of parton shower and matrix elments
are reviewed, which are then used in the description of the
merging. Section \ref{sec:algorithm} describes the various steps in
the algorithm and how to calculate all the terms needed. The algorithm
is implemented toghether with \ariadne\cite{Lonnblad:1992tz} and the
results are presented in section \ref{sec:results}. Finally in section
\ref{sec:conclusions} our conclusions are presented.

\section{Theory}
\label{sec:theory}

This section contains some basic properties of parton showers and matrix
elements, including both tree-level and one-loop matrix elements. After
the theoretical background has been established our algorithm for
merging one-loop matrix elements and parton shower is presented.

\subsection{Parton showers}

Emissions in the soft and collinear limit can be resummed to all orders
using a parton shower. This is done using only the dominant behavior in
this limit which is translated into an emission probablility. The other
main component is the assumption that the emissions can be ordered. To
make the results exclusive, the probability for emitting a parton also
includes the probability that no emission has occured at a higher scale,
which is known as a Sudakov form factor. The result is a formalism where
each emission can be considered individually and generated according to
a resonably simple probability distribution, which is ideal for computer
simulations.

The parton shower is a good approximation for emissions near the soft
and collinear limits and can give a nice description for a large range
of observables, since the approximation is valid for the bulk of the
cross section. There are important exceptions though, which occur
mainly when you have several hard partons emitted at wide angle.

Different shower models use different choices to specify the ordering
variable, the most common choices being transverse momentum, angle and
virtuality. To formulate the parton cross sections in a general way we
simply denote the ordering variable $\rho$. The emission probability is
the product of a splitting function, which is a function of the
emissions kinematics, and the stong coupling $\as$. Most showers use a
running $\as$ with a transverse momentum as the scale, which is not
necessarily equal to the ordering variable. However, here we assume that
the ordering variable is the same as the scale in $\as$ for
notiational convenience. The cross sections for the parton
multiplicities for a shower that has evolved down to the shower cutoff
($\rho_c$) can be written in the following way.

\begin{eqnarray}
  \label{eq:psexpansion}
  d\sigma_{0} &=& C\sup{PS}_0(\Omega\sup{PS}_0)\,\sud{0}(\rho_0,\rho_c)
    d\Omega\sup{PS}_0\nonumber\\
  d\sigma_{1} &=& C\sup{PS}_1(\Omega\sup{PS}_1)\,\as(\rho_1)
  \sud{0}(\rho_0,\rho_1)\sud{1}(\rho_1,\rho_c)d\Omega\sup{PS}_1 \nonumber\\
  d\sigma_{2} &=& C\sup{PS}_2(\Omega\sup{PS}_2)\,\as(\rho_1)\as(\rho_2)
  \sud{0}(\rho_0,\rho_1)\sud{1}(\rho_1,\rho_2)\sud{2}(\rho_2,\rho_c)
  d\Omega\sup{PS}_2\nonumber\\
  &\vdots&\nonumber\\
  d\sigma_{n} &=& C\sup{PS}_{n}(\Omega\sup{PS}_n)\sud{n}(\rho_n,\rho_c)
    \prod_{i=1}^n\as(\rho_i)
  \sud{i-1}(\rho_{i-1},\rho_i)
  d\Omega\sup{PS}_n\nonumber\\
  &\vdots&
\end{eqnarray}
The parton shower phase space is described by 
\begin{equation}
\Omega\sup{PS}_n=(\mathbf{q}_1,\ldots,\mathbf{q}_m, \rho_1,\vec{x}_1,
\rho_2,\vec{x}_2,\ldots,\rho_n,\vec{x}_n),
\end{equation}
where $\mathbf{q}$ denotes the momenta of the $m$ (usually two)
outgoing partons at Born level, $\rho$ is the
value of the ordering variable and $\vec{x}$ are the other kinematical
variables that describe each emission.  $\sud{n}(\rho_{n},\rho_{n+1})$
denotes the Sudakov form factor which is the probability that no
emission occurs from the $n$-parton state, $S_n$, between the scales
$\rho_n$ and $\rho_{n+1}$, and the $C\sup{PS}_n$-coefficients are the
Born-level matrix element multiplied with the products of splitting
functions in the shower and depend on all $\rho_i$ and $\vec{x}_i$
with $i\le n$.

The Sudakov form factors in the shower is an approximate way of
calculating the virtual diagrams to all orders. In the angular ordered
shower in \herwig this is done by an analytical calculation based on
the production scale of the various partons, but other showers, such
as the parton showers in \pythia and the dipole shower in \ariadne,
uses the actual parton state when calculating the Sudakov form factor
as an explicit no-emission probability. The algorithm described in
later section requires Sudakov form factors that factorize
($\sud{i}(\rho_1,\rho_2) \sud{i}(\rho_2,\rho_3) =
\sud{i}(\rho_1,\rho_3)$) which is the case if the Sudakov form factors
only depend on the intermediate state. In this paper the notation used
reflects the dependency on the parton state, which is denoted by a
subscript. The Sudakov form factors can explicitly be written

\begin{equation}
  \label{eq:sud}
  \sud{n}(\rho_i,\rho_{i+1})=\exp\left(-\int_{\rho_{i+1}}^{\rho_i}
    d\rho\as(\rho) \Gamma_{S_n}(\rho) \right),
\end{equation}
where $\Gamma$ denotes the branching probability for the specific parton
state.

It should be noted that the way the virtual corrections are
approximated in the shower means that the sum of all the parton cross
sections is equal to the Born cross section:

\begin{equation}
    \sum_{i=0}^\infty \sigma_i = \int C\sup{PS}_0(\Omega\sup{PS}_0)
    d\Omega\sup{PS}_0 = \sigma_{\mrm{Born}}.
\end{equation}

This means that the parton shower does not properly approximate the
higher-order corrections to the total cross section. Instead it is
common to include a $K$-factor by scaling all the cross sections with
the N$^l$LO cross section divided by the Born cross
section. However, it does nothing to improve the shape observables or
the relative abundance of different parton multiplicities.

Measurements of $\as$ have been done with better and better accuracy,
typically using precision data from LEP (see \eg\
\cite{Becher:2008cf}). If the same values of $\as$ would be used within
a parton shower it would not describe data well. The reason is that the
shower has a tendency to underestimate emission probabilities,
especially for hard emissions. The shower therefore gives a better
description of data if a higher value of $\as$ is used. To get the best
possible fit, the parton shower implementations usually make $\as$
tunable. This is frequently done by doing a one- or two-loop $\as$
evolution and making $\lqcd$ a parameter to be fit to data.

Making $\as$ tunable is equal to modifying the scale used in the
evaluation of $\as$, apart from corrections related to flavour
thresholds. This means that it is possible to use a different scale when
evaluating $\as$ and use the world average $\as$ from precision
measurements, which is illustrated in the equation below.
\begin{eqnarray}
\as\sup{PS}(\rho) &=&  \as\sup{WA}(b\rho)
\label{eq:bparam}
\end{eqnarray}
This is something that is used in later sections of this paper.

It is possible to expand the parton cross sections in the shower in
powers of $\as$. To do this the exponential in each Sudakov form
factors has to be expanded and the running of the coupling taken into
account.  The relative change of the cross section at each order of
$\as$ is denoted by $c_{n,m}$, where $n$ is the order in $\as$ of the
relevant tree diagram and $m$ is the number of loops. Using a
renormalization scale, $\mu$, the parton cross sections in
\eqref{eq:psexpansion} can be written as

\begin{eqnarray}
  \label{eq:psexpansion2}
  d\sigma_{0} &=& C\sup{PS}_0(\Omega\sup{PS}_0)\,
  (1+c_{0,1}\sup{PS}(\Omega\sup{PS}_0)\as(\mu)+
   c_{0,2}\sup{PS}(\Omega\sup{PS}_0,\mu)\as^2(\mu)+\ldots)
  d\Omega\sup{PS}_0 \nonumber\\
  d\sigma_{1} &=& C\sup{PS}_1(\Omega\sup{PS}_1)\,\as(\mu)
  (1+c_{1,1}\sup{PS}(\Omega\sup{PS}_1,\mu)\as(\mu)+
   c_{1,2}\sup{PS}(\Omega\sup{PS}_1,\mu)\as^2(\mu)+
   \ldots)d\Omega\sup{PS}_1 \nonumber\\
  d\sigma_{2} &=& C\sup{PS}_2(\Omega\sup{PS}_2)\,\as^2(\mu)
  (1+c_{2,1}\sup{PS}(\Omega\sup{PS}_2,\mu)\as(\mu)+
   c_{2,2}\sup{PS}(\Omega\sup{PS}_2,\mu)\as^2(\mu)+\ldots)
  d\Omega\sup{PS}_2\nonumber\\
  &\vdots&\nonumber\\
  d\sigma_{n} &=& C\sup{PS}_n(\Omega\sup{PS}_2)\as^n(\mu)
  (1+c_{n,1}\sup{PS}(\Omega\sup{PS}_n,\mu)\as(\mu)+
   c_{n,2}\sup{PS}(\Omega\sup{PS}_n,\mu)\as^2(\mu)+\ldots)
  d\Omega\sup{PS}_n\nonumber\\
  &\vdots&
\end{eqnarray}

All the higher order changes to the cross section (except $c_{0,1}$ in
case the Born level contains no powers of \as, which is assumed here)
have a dependence on the renormalization scale, when one expands in
terms of a fixed coupling constant. This happens because the shower
uses a running $\as$ and when changing the scale to $\mu$ there are
residual terms that needs to be absorbed into the parton shower
coefficients.  These effects are described further in later sections.

Note that the parton shower provides a value for the parton cross
sections that includes terms of all orders in $\as$, but the
coefficients are only approximately correct. The goal of merging
algorithms is to replace some of the coefficients by the exact results
in order to minimize the effects of the approximations done in the
parton shower. In the following we discuss how these terms can be
calculated and what kind of results one can achieve.

\subsection{Matrix elements}

Processes calculated through matrix elements means that one is
calculating the amplitudes of the Feynman diagrams directly. This is
easy to do for $2 \rightarrow 2$ processes, but gets increasingly
difficult for larger number of external legs or if one includes loops.

To calculate an observable using matrix elements is equivalent to
exactly calculating the terms in an $\as$ expansion one term at a
time.  The advantage of matrix elements is that they are exact up to
the calculated order. In certain regions of phase space this approach
works fine, but for collinear and soft emissions there are
divergencies in the matrix elements which prevents the $\as$ expansion
from converging.

The leading order term for final states with several outgoing partons
are tree-level matrix elements (no loops). If one expands the
differential cross sections in different multiplicities one arrives at
the following

\begin{eqnarray}
  \label{eq:meexpansion}
  d\sigma_{0} &=& C\sup{ME}_0(\Omega_0)\,
  d\Omega_0\nonumber\\
  d\sigma_{1} &=& C\sup{ME}_1(\Omega_1)\,\as
  d\Omega_1 \nonumber\\
  d\sigma_{2} &=& C\sup{ME}_2(\Omega_2)\,\as^2
  d\Omega_2 \nonumber\\
  &\vdots&\nonumber\\
  d\sigma_{n} &=& C\sup{ME}_n(\Omega_n)\,\as^n
  d\Omega_n \nonumber\\
  &\vdots&
\end{eqnarray}
where we have used the short-hand notation
\begin{equation}
\Omega_n=(\mathbf{p}_1,\ldots,\mathbf{p}_{n+m}).
\end{equation}
Here $\mathbf{p}$ is used to denote the momenta of the outgoing partons
in the matrix element. The Born-level diagram considered has $m$
outgoing partons and $n$ denotes the number of extra outgoing partons.
For simplicity we assume throughout that the Born-level cross sections
does not contain any powers of \as, but this requirement can easily be
relaxed.

The tree-level expansion is divergent and one needs to introduce a
phase-space cut to avoid collinear and soft configurations. Another
important property of tree-level matrix elements is that they describe
inclusive quantities. If one, \eg, integrates the three-parton matrix
element according to a jet definition, this yields the cross section
for a configuration with \emph{at least} three jets.

There a several methods for calculating tree-level matrix elements which
have been implemented as parts of automated programs. These programs
generate all the possible diagrams, sum them and generate events
accordingly. Examples of such programs include
\madevent \cite{Alwall:2007st} and \alpgen \cite{Mangano:2002ea}.

There are many uncertainties associated with tree-level matrix element
that can be better controlled if one could include the next order in
the $\as$ expansion, which would mean including one-loop matrix
elements.  One of the problems with loop matrix elements is that they
are infinite and frequently negative. Only the sum of the associated
real emission with the virtual one is a finite quantity. In practice
one assumes that emissions within a region are unresolved and that
their amplitude can be added to the virtual contribution. There are a
few choices to be made, depending on when to consider an emission
unresolved and how to map the unresolved contribution onto the
virtual one.

Most one-loop matrix elements are calculated with a method called
Catani--Seymour dipole subtraction \cite{Catani:1996jh, Catani:1996vz}.
This method uses a function calculated analytically from dipoles that is
added to the virtual contribution and subtracted from the real
contribution. This way of calculating one-loop matrix elements has been
proven to work quite well, but it needs to be modified to be applied in
to our algorithm. The reason is that for our matching algorithms to
work, a strict phase space cut is needed to separate resolved and
unresolved emissions. This can be accomplished by modifying the
subtraction scheme outside the singular regions.

If a jet cutoff, $\ycut$, is used to determine when an emission is
resolved, and the renormalization scale is set to $\mu$, one can
formulate the cross sections for one-loop matrix elements in the
following way.

\begin{eqnarray}
  \label{eq:1lmeexpansion}
  d\sigma_{0} &=& C\sup{ME}_0(\Omega_0) (1+\as(\mu)
  c\sup{ME}_{0,1}(\Omega_0,\ycut))\,d\Omega_0 \nonumber\\
  d\sigma_{1} &=& C\sup{ME}_1(\Omega_1)\,\as(\mu)
  (1+\as(\mu)c\sup{ME}_{1,1}(\Omega_1,\mu,\ycut))
  \Theta(y-\ycut)d\Omega_1 \nonumber\\
  d\sigma_{2} &=& C\sup{ME}_2(\Omega_2)\,\as^2(\mu)
  (1+\as(\mu) c\sup{ME}_{2,1}(\Omega_2,\mu,\ycut))
  \Theta(y-\ycut)d\Omega_2\nonumber\\
  &\vdots&\nonumber\\
  d\sigma_{n-1} &=& C\sup{ME}_{n-1}(\Omega_{n-1})\,\as^{n-1}(\mu)
  (1+\as(\mu) c\sup{ME}_{n-1,1}(\Omega_{n-1},\mu,\ycut))
  \Theta(y-\ycut) d\Omega_{n-1} \nonumber\\
  d\sigma_{n} &=& C\sup{ME}_{n}(\Omega_n)\,\as^n(\mu)
  \Theta(y-\ycut) d\Omega_n
\end{eqnarray}

When one-loop matrix elements are calculated, the unresolved parton of
the tree-level matrix element with one more outgoing particles needs to
be added. The resolved part is considered to have a higher multiplicity
but also needs to be included in the calculation. This means that the
highest multiplicity is always calculated to tree-level accuracy.

The renormalization scale used in the calculation enters not only as a
scale in $\as$, but also affects the one-loop terms (except for
$c\sup{ME}_{0,1}$ in case the Born level contains no powers of
\as). It is quite simple to see how the renormalization scale enter if
one considers the running coupling, which can be expanded as
\begin{eqnarray}
    \as(\mu') & = & \as(\mu)\left(1+\as(\mu)
        \frac{\log(\mu/\mu')}{\alpha_0} + {\cal O}(\as^2(\mu))\right)
        \nonumber\\
\alpha_0 &=& \frac{2\pi}{\beta_0} = \frac{6\pi}{33-2n_f}
\end{eqnarray}

This means that a change in the renormalization scale of the first
term leaves a remnant term in the next order in $\as$ simply through
the running, which is something that needs to be taken into account if
one varies the renormalization scale. The effect can be studied by
expanding $\as$ explicity for one of the multiplicities
\begin{eqnarray}
  && \as^{l}(\mu)
  (1+\as(\mu) c\sup{ME}_{l,1}(\Omega_{l},\mu,\ycut)) = 
  \nonumber\\
  && \as^{l}(\mu')
  (1+l \as(\mu') \frac{\log(\mu'/\mu)}{\alpha_0} + \as(\mu')
   c\sup{ME}_{l,1}(\Omega_{l},\mu,\ycut) + {\cal O}(\as^2(\mu'))),
\end{eqnarray}
which leads to the following scale dependence
\begin{eqnarray}
  c\sup{ME}_{l,1}(\Omega_{l},\mu',\ycut)) = 
  c\sup{ME}_{l,1}(\Omega_{l},\mu,\ycut) +
  l\frac{\log(\mu'/\mu)}{\alpha_0} + {\cal O}(\as(\mu'))).
\label{eq:cmmodmu}
\end{eqnarray}

The renormalization scale can also be used to tune the
different jet fractions, which is known as optimized perturbation
theory\cite{Stevenson:1981vj}. Going beyond one-loop matrix elements
or going to processes with incoming hadrons, there is also a
renormalization scheme dependence to be considered.

\subsection{Merging parton showers and tree-level matrix elements}

The purpose of merging algorithms is to improve the description of jet
observables without changing things such as the internal jet structure
which is described well by the shower. At a more formal level the goal
is to replace some coefficients in the expansion of the parton shower
with their correct counterparts from the matrix elements.

There are several algorithms formulated with this purpose in mind. The
main ones are CKKW\cite{Catani:2001cc, Krauss:2002up},
CKKW-L\cite{Lonnblad:2001iq, Lavesson:2005xu}, MLM \cite{MLM,
  Mangano:2006rw} and Pseudo-Shower \cite{Mrenna:2003if}. Their
advantages and disadvantages when applied to e$^+$e$^-$ annihilation
is discussed thoroughly in \cite{Lavesson:2007uu}. Here the discussion
is limited to how the CKKW-L algorithm can be extended to also include
one-loop matrix elements, but most of the general ideas can be applied
also to CKKW.

Before going to one-loop matrix element, the mechanism used to merge
tree-level matrix elements needs to be understood. With tree-level
matrix elements two issues need to be resolved, namely to divide the
phase space for emissions between the parton shower and the matrix
element and to introduce Sudakov form factors to make the matrix
elements exclusive. Essentially one would like to replace the product
of splitting functions present in the cross sections for various
processes in the parton shower with the correct tree-level matrix
element.

The phase space is to be divided in such a way that the region for
allowed emissions from the matrix element and the parton shower cover
the entire phase space with no overlaps. Failing to do this
consistently results in double counting or dead regions. The scale
that defines the border between the matrix-element and the
parton-shower phase space is known as the merging scale, and is
usually defined using a jet clustering algorithm.

The Sudakov form factors are introduced by using a constructed shower
history, which is done by considering all possible shower histories
for the states generated according to the matrix element and selecting
one with a probability proportional to the product of the
corresponding splitting functions in the shower. The actual shower is
then used for calculating the Sudakov form factors, which has the
advantage that any non-leading effects that were introduced in the
shower are also included. For further details on this procedure we
refer the reader to \cite{Lonnblad:2001iq,Lavesson:2005xu}.

A particular shower scenario is dependent on the emission scales of
the shower (denoted $\rho$) and other shower variables such as energy
fraction and angular orientation, which are denoted simply by
$\vec{x}$.  A complete set of scales and other variables can be used
to yield a shower history composed of specific states, denoted
$S_i$. Looking at the differential exclusive cross section for $n$
emitted partons, the parton shower yields the following

\begin{eqnarray}
  \frac{d\sigma\sup{PS}_n}{d\Omega\sup{PS}_n}&=&
  K C_n\sup{PS}(\Omega\sup{PS}_n)
  \Delta_{S_n}(\rho_n,\rho_c)
  \prod_{i=1}^{n}\as(b\rho_i) \Delta_{S_{i-1}}(\rho_{i-1},\rho_i),
    \label{eq:shxsec}
\end{eqnarray}
where $\rho_0$ is the maximum scale, $\rho_c$ is the shower cutoff scale
and $b$ is the parameter introduced in equation (\ref{eq:bparam}). An overall
N$^l$LO $K$-factor, $K=1+\sum_{i=1}^l k_i\as^i(\mu)$, has also been
included.

The above expression is to be compared with the appropriate tree-level
matrix element. The tree-level matrix element is an inclusive quantity
and do not cover the full phase space, which is denoted by including a
step function with the matrix element cutoff which is set equal to the
merging scale, $\yms$. The cross section for the tree-level matrix
element can be formulated in the following way.

\begin{eqnarray}
  \frac{d\sigma\sup{ME}_n(\yms)}{d\Omega_n}&=&
  C_n\sup{ME}(\Omega_n)\as^{n}(\mu)
  \Theta(y(\Omega_n)-\yms)
\end{eqnarray}

By selecting one history out of all possible histories, the matrix
element can be mapped onto the shower phase space formulation, which can
be described by the mapping $\Omega_n\mapsto\Omega\sup{PS}_n$. The
matrix element cross section can be written in terms of parton shower
phase space in the following way.

\begin{eqnarray}
  \frac{d\sigma\sup{ME}_n(\yms)}{d\Omega\sup{PS}_n}&=&
  C_n\sup{ME}(\Omega\sup{PS}_n)\as^{n}(\mu)
  \Theta(y(S_n)-\yms)
\end{eqnarray}

The Sudakov form factors are then introduced in the same way as in
equation (\ref{eq:shxsec}). In addition, the coupling constant is
reweighted to use the emission scales instead of a fixed
renormalization scale. The procedure results in the following
exclusive cross section.

\begin{eqnarray}
  \frac{d\sigma\sup{}_n(\yms)}{d\Omega\sup{PS}_n}&=&
  K C_n\sup{ME}(\Omega\sup{PS}_n)\as^{n}(\mu)\Theta(y(S_n)-\yms)
  \Delta_{S_n}(\rho_n,\rho_c) \times\nonumber\\
  &&\prod_{i=1}^n
  \frac{\as(b\rho_i)}{\as(\mu)} \Delta_{S_{i-1}}(\rho_{i-1},\rho_i)
\label{eq:mepstl}
\end{eqnarray}

This expression is fully exclusive in the same way as a state generated
by the shower. However, the tree-level matrix element is only allowed to
generate emissions above the merging scale. The scheme therefore needs
to be supplemented by introducing a way of allowing the shower to
generate extra emissions below the merging scale and the methods for
accomplishing this is the subject of the next section.

\subsection{Adding parton showers to multi-parton states}
\label{sec:adding-part-show}

If the merging scale, $\yms$, is defined in the same way as the parton
shower ordering variable, the adding of a parton shower is fairly
trivial. For most parton showers you can simply shower each parton
individually and use $\yms$ as the maximum scale for the ordering
variable.

If, however, the merging scale and ordering scale are different, \eg\
the merging scale is defined in invariant mass, while the shower
ordering is in transverse momentum (as is the case in our studies
below in section \ref{sec:results}), the problem becomes non-trivial.

In the original CKKW formulation, the problem was solved by
introducing the concept of a ``vetoed shower''. Here, each parton is
allowed to shower, starting from a value of the ordering variable
typically given by the maximum possible scale in the Born-level
process. Each emission is then checked so that it if is above the
merging scale, $\yms$, the emission is discarded, allowing the shower
to continue to evolve down to lower evolution scales.

There is, however, a problem with this procedure, as was noted in
\cite{Nason:2004rx} and \cite{Lavesson:2007uu}, related to the fact
that the shower may be allowed to make effectively unordered emissions.
To understand the problem, we consider a partonic state corresponding to
$n$ parton emissions beyond the Born level. The state can be mapped onto
a set of intermediate states and scales, where $\rho_n$ represents the
scale of the last emission. Now if the merging scale is very different
from the ordering scale, it may very well happen in a vetoed shower that
an emission with $y<\yms$ and $\rho>\rho_n$ is generated, which breaks
the ordering, since in the shower such an emission should have been
emitted from an intermediate state. Breaking the ordering results in the
wrong colour structure, which may result in incorrect treatment of
coherence effect, and different kinematics, which may give unwanted
suppressions in some regions of phase space.

So far, two solutions to this problem has been presented. The CKKW-L
approach and the so-called ``truncated'' vetoed shower. Both
approaches requires that not only emission scales are reconstructed as
in CKKW, but also the full kinematics of the complete shower history
with on-shell intermediate states, $S_i$.

The truncated shower \cite{Nason:2004rx}\footnote{Here we only give our
rough interpretation of the truncated shower. For a more detailed
description we refer to \cite{Nason:2004rx}} is way of allowing the
shower to generate emissions from the intermediate states in the shower
history and thereby preserve the ordering. The vetoed shower is started
from the Born-level state, $S_0$ with the corresponding maximum scale
$\rho_0$ and is vetoed in the same way as above. When the shower
evolution comes down to $\rho_1$, it is stopped and the reconstructed
emission $(\rho_1,\vec{x}_1)$ that was generated by the matrix element
is inserted by hand. The vetoed evolution is then continued down to
$\rho_2$, where the next reconstructed emission is inserted, and so on,
in a way such that the kinematics of the partons in the original state
is minimally disturbed.

The philosophy of the CKKW-L approach is quite different, in that
unordered emissions are simply forbidden. This means that as soon as
there is one emission below the merging scale, that emission and all
subsequent emission (above and below the merging scale) are generated by
the shower. When adding a shower to a $n$-parton state with the
reconstructed scale $\rho_n$, the shower is started from $\rho_n$ and
the first emission is forced to be below the merging scale, but later
emissions have no such restrictions. The reweighting for the same state
is done by using the full Sudakov form factor down to the scale
$\rho_n$. The state which are thus forbidden can be generated by matrix
elements for lower multiplicity together with the shower.

The CKKW-L approach can be applied to the cross section in equation
(\ref{eq:mepstl}) modifying the last Sudakov form factor and adding a
shower. The first emission is added with the restriction that it should
be below $\rho_n$ and the merging scale. The emission probability can be
described as 
\begin{eqnarray}
dP &=& \as(b\rho) \Gamma_{S_n}(\rho,\vec{x})
\Theta(\yms-y(S_n,\rho,\vec{x})) \Delta_{S_n}(\rho_n,\rho;<\yms) d\rho
d\vec{x},
\end{eqnarray}
where
\begin{eqnarray}
  \Delta_{S_n}(\rho_n,\rho;<\yms) =
  \exp\left(-\int_{\rho}^{\rho_n}d\rho' d\vec{x}
  \as(b\rho') \Gamma_{S_i}(\rho',\vec{x})
  \Theta(\yms-y(S_n,\rho',\vec{x}))\right).
  \label{eq:modsud}
\end{eqnarray}
In addition to the probability described above the CKKW-L procedure
specifies that one should also include the Sudakov form factor between
the generated emission $\rho_{n+1}$ and the previous emission $\rho_n$
and above the merging scale,
$\Delta_{S_n}(\rho_n,\rho_{n+1};>\yms)$. The cascade is then continued
from $\rho_{n+1}$ with no veto.

An alternative way of performing the same calculation is to generate one
emission starting from $\rho_n$ and discard the entire event if the
emission was above the merging scale. It is equivalent to the above
procedure since the probability for discarding the event is equal to the
Sudakov form factor calculated at the end. This is the way the algorithm
was formulated in \cite{Lonnblad:2001iq}

The exclusive cross section in equation (\ref{eq:mepstl}) should be
modified accordingly. Assuming that one emission has been added using the
procedure describe above, one can formulate the cross section in the
following way.
\begin{eqnarray}
  \frac{d\sigma\sup{}_n(\yms)}{d\Omega\sup{PS}_n}&=&
  K C_n\sup{ME}(\Omega\sup{PS}_n)\as^{n}(\mu)\Theta(y(S_n)-\yms)
  \Delta_{S_n}(\rho_n,\rho_{n+1};>\yms) \times\nonumber\\
  &&\prod_{i=1}^n
  \frac{\as(b\rho_i)}{\as(\mu)} \Delta_{S_{i-1}}(\rho_{i-1},\rho_i)
\label{eq:mepsinc}
\end{eqnarray}

Another important issue to consider when adding a parton shower is how
to handle the highest multiplicity states with $n=N$. Clearly we must
here not veto emissions above $\yms$, since this would artificially
suppress final states with more than $N$ emissions above this
scale. In this case no extra Sudakov form factors needs to be included
and the cross section is given by
\begin{eqnarray}
  \frac{d\sigma\sup{}_n(\yms)}{d\Omega\sup{PS}_n}&=&
  K C_n\sup{ME}(\Omega\sup{PS}_n)\as^{n}(\mu)\Theta(y(S_n)-\yms)
  \prod_{i=1}^n
  \frac{\as(b\rho_i)}{\as(\mu)} \Delta_{S_{i-1}}(\rho_{i-1},\rho_i).
\label{eq:mepsinchmult}
\end{eqnarray}
The only constraint on the shower is that the first emission should be
below $\rho_n$.

In the CKKW-L algorithm, emissions are thus corrected with the full
matrix element only if they are among the $N$ hardest (according to
the parton shower ordering) \textit{and} are all above the merging
scale.

\subsection{Extending to one-loop MEs}
\label{sec:extending-one-loop}

To be able to extend the algorithm to also include one-loop matrix
elements a new set of issues has to be addressed. First and foremost,
the one-loop matrix element contains a terms which is one order higher
in \as. To be able to apply a correction, the shower cross section,
\eqref{eq:shxsec}, therefore needs to be expanded to that level, with
a fixed renormalization scale $\mu$. The $\as$ expansion of the Sudakov
form factor and the running coupling can be written
\begin{eqnarray}
  \Delta_{S_i}(\rho_i,\rho_{i+1})&=&
  1-\int_{\rho_{i+1}}^{\rho_i}d\rho\as(b\rho)
  \Gamma_{S_i}\sup{PS}(\rho) + \ldots \nonumber\\
  &=& 1-\as(\mu)\int^{\rho_i}_{\rho_{i+1}}d\rho\Gamma_{S_i}\sup{PS}(\rho)
   + {\cal O}(\as^2(\mu))
    \label{eq:sudexpand} \\
  \as(b\rho)&=&\as(\mu)\left(1 + \as(\mu)\frac{\log(\mu/(b\rho))}{\alpha_0}
     + {\cal O}(\as^2(\mu))\right) \label{eq:asexpand},
\end{eqnarray}
where the possibility of modifying the scale used in $\as$ in the shower
has been included from equation (\ref{eq:bparam}).

This means the parton shower cross section in \eqref{eq:shxsec} can be
rewritten as
\begin{eqnarray}
  \frac{d\sigma\sup{PS}_n}{d\Omega\sup{PS}_n}&=&
  C_n\sup{PS}(\Omega\sup{PS}_n)\as^{n}(\mu)
  \left[1+
    \as(\mu)\left\{k_1+\sum_{i=1}^n\frac{\log(\mu/(b\rho_i))}{\alpha_0}
    \right.\right.\nonumber\\
    & &\left.\left.\qquad\qquad\qquad-\sum_{i=0}^{n-1}
      \int_{\rho_{i+1}}^{\rho_i}d\rho\Gamma_{S_i}(\rho)
    -\int_{\rho_c}^{\rho_n}d\rho\Gamma_{S_n}(\rho)\right\}
    + {\cal O}(\as^2(\mu))\right].\label{eq:PSexpand}
\end{eqnarray}
Note that the extra term that appears because of the running coupling is
equivalent to the term which appears if the renormalization scale of the
one-loop matrix element is changed, derived explicity in equation
(\ref{eq:cmmodmu}).

The one-loop matrix elements need to be put through the procedure
described earlier to construct a shower history. The procedure yields
the following form for the cross section.

\begin{eqnarray}
  \frac{d\sigma\sup{ME}_n(\yms)}{d\Omega\sup{PS}_n}&=&
  C_n\sup{ME}(\Omega\sup{PS}_n)\as^{n}(\mu)
  \left[1+\as(\mu)c_{n,1}\sup{ME}(\Omega\sup{PS}_n,\mu,\yms)\right]
  \Theta(y(S_n)-\yms)
\end{eqnarray}

The cross section contains the same terms as the tree-level cross
section plus the next order correction to the matrix element, which is a
sum of the virtual diagrams of multiplicity $n$ and the real diagrams
with multiplicity $n+1$ kinematics below the merging scale $\yms$.

If the merging scale is defined using the same scale as the ordering variable
in the shower, no further modifications to the matrix element would have
been necessary. However, to preserve the ordering of the shower, extra
Sudakov form factors need to be included. The phase space in question is
where emissions have a higher ordering variable ($\rho$) than the last
emission of the matrix element ($\rho_n$) and are below the merging
scale. These emissions would violate the shower ordering and are
therefore forbidden, but one still needs to include virtual corrections
in this region, which is done using Sudakov form factors. The reweighted
one-loop cross sections can be written as
\begin{eqnarray}
  \frac{d\sigma\sup{}_n(\yms)}{d\Omega\sup{PS}_n}&=&
  C_n\sup{ME}(\Omega\sup{PS}_n)\as^{n}(\mu)
  \left[1+\as(\mu)c_{n,1}\sup{ME}(\Omega\sup{PS}_n,\mu,\yms)\right]
  \Theta(y(S_n)-\yms)\times\nonumber\\
  &&\prod_{i=0}^{n-1} \Delta_{S_i}(\rho_i,\rho_{i+1};<\yms),
    \label{eq:1loopSud}
\end{eqnarray}
where the definition for the Sudakov from factor was presented in
equation (\ref{eq:modsud}). Note that no reweighting is necessary for
the lowest multiplicity processes, since if there are no emissions in
the matrix element state, there are no regions of phase space where the
shower can generate emissions violating the ordering requirement.

The other component that goes into the merging is the shower where all
the terms corresponding to the one-loop matrix element have been
subtracted. When doing the subtraction we choose to work with a cascade
which is already corrected with tree-level matrix elements. The idea is
to reweight the tree-level matrix element the same way as CKKW-L, which
is described by equation (\ref{eq:mepsinc}), and then subtract the terms
corresponding to the one-loop matrix element. Note that one of the
Sudakov form factors has a dependency on the scale of the next emission
performed by the shower $\rho_{n+1}$, which was described in section
\ref{sec:adding-part-show}.

The terms of order $\as^{n+1}$ in equation (\ref{eq:PSexpand}) needs
to modified to comply with the phase space restrictions of the
one-loop matrix element. The shower is formulated in a way where
emissions can be anywhere within the allowed shower phase space,
whereas the matrix element is restricted to emissions above the
merging scale. This difference in phase space does not affect the
expansion of the running coupling, but must be included in the
integration over the branching probability. The full formula including
tree-level matrix element corrected with CKKW-L and the subtraction of
the one-loop terms is the following.

\begin{eqnarray}
  \frac{d\sigma\sup{PScorr}_{n}(\yms)}{d\Omega\sup{PS}_n}&=&
  C_n\sup{ME}(\Omega\sup{PS}_n)\as^{n}(\mu) \times\nonumber\\
  & & \left[K\Delta_{S_n}(\rho_n,\rho_{n+1};>\yms)\prod_{i=1}^n
  \frac{\as(b\rho_i)}{\as(\mu)} \Delta_{S_{i-1}}(\rho_{i-1},\rho_i)\right.
  \nonumber\\
  & &\left.\qquad-\prod_{i=0}^{n-1}
    \Delta_{S_i}(\rho_i,\rho_{i+1};<\yms)
    \left\{1+k_1\as(\mu)+
    \as(\mu)\sum_{i=1}^n\frac{\log(\mu/(b\rho_i))}{\alpha_0}
    \right.\right.\nonumber\\
    & &\qquad\qquad-\as(\mu)\sum_{i=0}^{n-1}
      \int_{\rho_{i+1}}^{\rho_i}d\rho d\vec{x}\,\Gamma_{S_i}(\rho,\vec{x})
    \Theta(y(S_i,\rho,\vec{x})-\yms)
    \nonumber\\
    & &\qquad\qquad-\as(\mu)
      \int_{\rho_c}^{\rho_n}d\rho d\vec{x}\,\Gamma_{S_n}(\rho,\vec{x})
    \Theta(y(S_i,\rho,\vec{x})-\yms)\Bigg\}
  \Bigg]
    \label{eq:pscorr}
\end{eqnarray}

The samples described by \eqref{eq:1loopSud} and \eqref{eq:pscorr} are
added together in the end to form the parton multiplicity cross section
for one-loop matrix elements merged together with parton showers. When
implementing the algorithms one also allows for extra emissions
generated by the shower with the requirement that the first such
emission should have a lower scale than $\rho_n$ and be below the
merging scale. The details of the entire procedure is described in
section \ref{sec:algorithm}.

\subsection{Two partons at two loops}
\label{sec:twoloop}

In the case of $\ee\to$ hadrons, there is an additional matrix element
that can be included without considering higher order terms in the
running of $\as$, which is the two-loop matrix element for two
partons.  The reason is that the leading-order term for the two-parton
cross section does not include $\as$. The matrix element is also
fairly easy to simulate if one knows the other components at order
$\as^2$. Working with two-loop terms means that the equation
(\ref{eq:sudexpand}) needs to be expanded one more order including the
parameter $b$, defined in equation (\ref{eq:bparam}). The expanded
Sudakov form factor can be written in the following way.
\begin{eqnarray}
  \Delta_{S_i}(\rho_i,\rho_{i+1})&=&
  1-\int_{\rho_{i+1}}^{\rho_i}d\rho\as(b\rho) \Gamma_{S_i}\sup{PS}(\rho)
+\frac{1}{2!} \left(\int_{\rho_{i+1}}^{\rho_i}d\rho\as(b\rho)
\Gamma_{S_i}\sup{PS}(\rho)\right)^2 - \ldots \nonumber \\
  &=& 1-\as(\mu)\int^{\rho_i}_{\rho_{i+1}}d\rho\Gamma_{S_i}\sup{PS}(\rho)
   -\as^2(\mu)\int^{\rho_i}_{\rho_{i+1}}d\rho\frac{\log(\mu/(b\rho))}{\alpha_0}
    \Gamma_{S_i}\sup{PS}(\rho) + \nonumber\\
    &&\frac{1}{2} \as^2(\mu)\left(\int^{\rho_i}_{\rho_{i+1}}d\rho
     \Gamma_{S_i}\sup{PS}(\rho)\right)^2
    + {\cal O}(\as^3(\mu))
\end{eqnarray}
Apart from the terms above, the $K$-factor of order $\as^2$ needs to be
included, but for the running $\as$ it is sufficient to include the
first order expansion of $\as$ given by equation (\ref{eq:asexpand}).
The tree-level matrix element can be modified in the following fashion,
which is to be added to the event generated according to the two loop
matrix element.
\begin{eqnarray}
  \frac{d\sigma\sup{PScorr}_2(\yms)}{d\Omega\sup{PS}_0}&=&
  C_0\sup{ME}(\Omega\sup{PS}_0)
  \left[K\Delta_{S_0}(\rho_0,\rho_1;>\yms)
  -\Bigg\{1+k_1\as(\mu)+k_2\as^2(\mu)
    \right.\nonumber\\
    & &\left.\left.\qquad\qquad-\as(\mu)(1+k_1\as(\mu))
      \int_{\rho_c}^{\rho_0}d\rho d\vec{x}\,\Gamma_{S_0}(\rho,\vec{x})
    \Theta(y(S_i,\rho,\vec{x})-\yms)  
    \right.\right.\nonumber\\&& \left.\left.
    \qquad\qquad-\as^2(\mu)\int_{\rho_c}^{\rho_0}d\rho d\vec{x}\,
    \frac{\log(\mu/\rho)}{\alpha_0} \Gamma_{S_0}(\rho,\vec{x})
    \Theta(y(S_i,\rho,\vec{x})-\yms)  
    \right.\right.\nonumber\\&& \left.\left.
    \qquad\qquad+\frac{\as^2(\mu)}{2}\left(\int_{\rho_c}^{\rho_0}d\rho
    d\vec{x}\, \Gamma_{S_0}(\rho,\vec{x})
    \Theta(y(S_i,\rho,\vec{x})-\yms)\right)^2\right\}
  \right]
    \label{eq:pscorr2}
\end{eqnarray}

This term represents the modified parton shower and added to the
two-parton matrix element calculated at two loops, the cascade
becomes corrected at one order higher in $\as$. Note that there is a
dependence on the scale of the next emission generated by the shower,
in this case $\rho_1$, in the same way as in equations
(\ref{eq:mepsinc}) and (\ref{eq:pscorr}).

The term described above should be added to the two-parton matrix
element calculated at two-loop accuracy. Since the two-loop matrix
element considered has the lowest multiplicity, no reweighting due to
the ordering requirement is necessary. If two-loop matrix elements with
higher parton multiplicity were to be included, they would have to be
reweighted in the same way as the one-loop matrix elements, described in
\eqref{eq:1loopSud}.

\section{The algorithm}
\label{sec:algorithm}

This section describes all the necessary steps needed to generate the
actual events. First the methods for calculating each individual
weight is presented and then each step in the algorithm is described.

\subsection{Calculating the terms}
\label{sec:algorithm-calc-terms}

Each of the weights that need to be calculated consists of a number of
common elements. This section describes how to calculate each individual
piece. All the terms are calculated similarly to the CKKW-L approach,
which means that the actual shower is used. However, it should be noted
that it is possible to do calculations along the same lines using the
analytical weights in the CKKW algorithm. Using the actual shower puts a
constraint on which parton showers can be used. The algorithm can only
be applied to showers with well defined intermediate states and Sudakov
form factors that factorizes, which is the same requirement as the
CKKW-L algorithm.

The weights described in this section are calculated using Monte Carlo
techniques, which means that there is an element of randomness in each
weight. The calculations presented give the correct value for each
weight if an average value is calculated. In the general framework
of Monte Carlo event generators this does not present a problem.

The simplest weight to be calculated is the plain Sudakov form factor,
denoted $\Delta_{S_i}(\rho_i,\rho_{i+1})$. This is done by generating
one emission with the shower starting from the state $S_i$ and the scale
$\rho_i$. If the emission is above the scale $\rho_{i+1}$ set the weight
to zero otherwise set it equal to one. This gives the correct behavior
since by definition the no-emission probability of the shower is equal
to the Sudakov form factor.

The next weight to be described is the Sudakov form factor that is
modified to exclude emissions above the merging scale denoted
$\Delta_{S_i}(\rho_i,\rho_{i+1};<\yms)$ and defined in equation
(\ref{eq:modsud}), and is used to reweight the states generated
according to one-loop matrix elements. In this case two different
scales are mixed and the scheme for the plain Sudakov form factor
needs to be supplemented with an accept/reject scheme. The calculation
is described in the following steps.
\begin{enumerate}
\item Feed the state $S_i$ into the shower and generate one emissions
starting from a scale $\rho_i$. This yields an emission scale $\rho$ and
a new state $S$.
\item Depending on the emission there are three options.
\label{step:modsud}
\begin{itemize}
\item If the scale of the generated emission $\rho$ is above
  $\rho_{i+1}$ and if the emission is above the merging scale ($y(S)
  > \yms$) generate a new emission from the state $S_i$, but this time
  using $\rho$ as the starting scale. Repeat step \ref{step:modsud}.
\item If the scale of the generated emission $\rho$ is below
  $\rho_{i+1}$ set the weight to one.
\item If the scale of the generated emission $\rho$ is above
  $\rho_{i+1}$ and if the emission is below the merging scale ($y(S)
  < \yms$) set the weight to zero.
\end{itemize}
\end{enumerate}

The only other terms that need to be calculated are the integrated
branching probabilities. First the calculation of the term $\as(\mu)
\int_{\rho_{i+1}}^{\rho_i}d\rho d\vec{x}\,\Gamma_{S_i}(\rho,\vec{x})
\Theta(y(S_i,\rho,\vec{x})-\yms)$ is described and later it is shown how
to extend the calculation to the other related terms. The calculation is
done by generating emissions in the entire avaliable phase space and
counting the total number. The procedure is described in the following
steps.
\begin{enumerate}
\item Use a fixed value for $\as=\as(\mu)$ (usually avaliable as an
  option in a parton shower program) and generate one emission from the
  state $S_i$ with a starting scale $\rho_i$. This yields an emission
  scale $\rho$ and a new state $S$.
\item If the emission has a scale $\rho > \rho_{i+1}$ and is above the
  merging scale $y(S) > \yms$ count the emission.
  \label{step:count}
\item Depending on the scale of the emission stop the algorithm or
  generate another emission.
  \label{step:continue}
  \begin{itemize}
  \item If the scale of the generated emission $\rho$ is above
    $\rho_{i+1}$ generate a new emission from the state $S_i$, but this time
    using $\rho$ as the starting scale. Repeat step \ref{step:count} and
    \ref{step:continue}.
  \item If the scale of the generated emission $\rho$ is below
    $\rho_{i+1}$ set the weight to the total number of emission counted in
    step \ref{step:count}.
  \end{itemize}
\end{enumerate}
The average number of emissions in the algorithm above gives the correct
value for the integral. To show that this is the case consider the
probability of $n$ emissions.
\begin{equation}
    P(n) = \Delta_{S_i}(\rho_i,\rho_{i+1};>\yms) \frac{1}{n!}
    \left(\as(\mu) \int_{\rho_{i+1}}^{\rho_i}d\rho
    d\vec{x}\,\Gamma_{S_i}(\rho,\vec{x})
    \Theta(y(S_i,\rho,\vec{x})-\yms)\right)^{n}.
\end{equation}
The average number of emission can be written as
\begin{eqnarray}
    \sum_{n=0}^{\infty} n P(n) &=& \Delta_{S_i}(\rho_i,\rho_{i+1};>\yms)
    \as(\mu)\int_{\rho_{i+1}}^{\rho_i}d\rho d\vec{x}\,
    \Gamma_{S_i}(\rho,\vec{x})
    \Theta(y(S_i,\rho,\vec{x})-\yms) \times\nonumber\\
    && \sum_{n=1}^{\infty} \frac{1}{(n-1)!}
    \left(\as(\mu)\int_{\rho_{i+1}}^{\rho_i}d\rho
    d\vec{x}\,\Gamma_{S_i}(\rho,\vec{x})
    \Theta(y(S_i,\rho,\vec{x})-\yms)\right)^{n-1} \nonumber\\
    &=& \as(\mu)\int_{\rho_{i+1}}^{\rho_i}d\rho
    d\vec{x}\,\Gamma_{S_i}(\rho,\vec{x})
    \Theta(y(S_i,\rho,\vec{x})-\yms).
\label{eq:pn}
\end{eqnarray}

The same algorithm can also be used to calculate the higher order terms
present in equation (\ref{eq:pscorr2}). The integrated branching
probability squared can be calculated by taking setting the weight to
$n(n-1)$, where $n$ is the number of emissions, which gives the correct
average since
\begin{eqnarray}
    \sum_{n=0}^{\infty} n (n-1) P(n) &=& \left(\as(\mu)
\int_{\rho_{i+1}}^{\rho_i}d\rho d\vec{x}\,\Gamma_{S_i}(\rho,\vec{x})
\Theta(y(S_i,\rho,\vec{x})-\yms)\right)^2.
\end{eqnarray}

There is one more term to be considered in equation (\ref{eq:pscorr2}),
which is 
\begin{equation}
\as^2(\mu)\int_{\rho_{i+1}}^{\rho_i}d\rho d\vec{x}\,
\frac{\log(\mu/(b\rho))}{\alpha_0} \Gamma_{S_i}(\rho,\vec{x})
    \Theta(y(S_i,\rho,\vec{x})-\yms).
\end{equation}
The way this integral is calculated is by an accept/reject scheme. The
first step is to rewrite the term where the value of the logarithm has
been divided by its maximum, which occurs for the minimum possible
value of $\rho$.
\begin{equation}
\left(\frac{\as(\mu)\log(\mu/(b\rho_{i+1}))}{\alpha_0}\right)
\left(\as(\mu)\int_{\rho_{i+1}}^{\rho_i}d\rho d\vec{x}\,
\frac{\log(\mu/(b\rho))}{\log(\mu/(b\rho_{i+1}))}
\Gamma_{S_i}(\rho,\vec{x}) \Theta(y(S_i,\rho,\vec{x})-\yms) \right)
\end{equation}
The factor to the right can now be simulated with a scheme simular to
the one described earlier. The only addition is that emissions are only
counted if $\log(\mu/(b\rho))/\log(\mu/(b\rho_{i+1}))>R$, where $R$ is a
random number between 0 and 1.

\subsection{The steps}
\label{sec:steps}

To generate the actual events we start with two different samples. One
is generated according to the one-loop matrix element and one generated
with tree-level matrix elements, where both samples are generated using
the same cutoff, $\yms$. The two samples are generated for all
multiplicates except for the highest one, where only the tree-level
matrix element is used. Different weights are calculated for the
events depending on if they were generated according to a one-loop
matrix element or a tree-level matrix element. Event are generated
according to the following steps:

\begin{enumerate}
\item Choose a merging scale $\yms$ and use the same scale as matrix
  element cutoff. Calculate the cross section for the one-loop matrix
  element with multiplicities $n<N$ and for tree-level matrix elements
  with multiplicities $n\le N$. Choose a matrix element with a
  probability proportional to its cross section.
\item Generate an event with a kinematic distribution in
  accordance with the chosen matrix element.
\item Construct a shower history by considering all possible histories
  and selection one with a probability proportional to the
  corresponding product of splitting functions. This leads to a set of
  states $S_{n} \ldots S_{0}$ and scales $\rho_{n} \ldots \rho_{1}$.
\item When generating the first emission from the shower, there are
  two cases to be considered.
\label{st:emission}
\begin{itemize}
\item If the event was generated according to a one-loop matrix
  element or according to a tree-level matrix element with a
  multiplicity less than the maximum ($n<N$), generate one emission
  starting from the state $S_n$ with a starting scale $\rho_n$, but veto
  any emission which is above the merging scale $\yms$
\item If the event was generated according to a tree-level matrix
  element and had the highest multiplicity ($n=N$), generate one
  emission from the state $S_n$ with a starting scale $\rho_n$.
\end{itemize}
\item The events are reweighted depending on type:
\begin{itemize}
\item If the event was generated according to a one-loop matrix
  element, reweight the event with a factor $\prod_{i=0}^{n-1}
  \Delta_{S_i}(\rho_i,\rho_{i+1};<\yms)$ according to the steps in
  subsection \ref{sec:algorithm-calc-terms}.
\item If the event was generated according to a tree-level matrix
  element, but did not have the highest multiplicity ($n<N$), then the
  the weight depends on the scale of the emission in step
  \ref{st:emission}, $\rho_{n+1}$. (If the shower cutoff was reached
  and no emission generated, set $\rho_{n+1} = \rho_c$.)  Reweight the
  event with
  \begin{eqnarray}
    &&K\Delta_{S_n}(\rho_n,\rho_{n+1};>\yms)\prod_{i=1}^n
    \frac{\as(b\rho_i)}{\as(\mu)} \Delta_{S_{i-1}}(\rho_{i-1},\rho_i)
    \nonumber\\
    &&-\prod_{i=0}^{n-1} \Delta_{S_i}(\rho_i,\rho_{i+1};<\yms)
    \left\{1+k_1\as(\mu)+
      \as(\mu)\sum_{i=1}^n\frac{\log(\mu/(b\rho_i))}{\alpha_0}\right.
    \nonumber\\
    &&\qquad\qquad\qquad\quad-\as(\mu)\sum_{i=0}^{n-1}
      \int_{\rho_{i+1}}^{\rho_i}d\rho d\vec{x}\,
      \Gamma_{S_i}(\rho,\vec{x})
      \Theta(y(S_i,\rho,\vec{x})-\yms)\nonumber\\
    &&\qquad\qquad\qquad\quad-\as(\mu)
      \int_{\rho_c}^{\rho_n}d\rho d\vec{x}\,\Gamma_{S_i}(\rho,\vec{x})
      \Theta(y(S_i,\rho,\vec{x})-\yms)\Bigg\},
  \end{eqnarray}
  according to the steps in subsection \ref{sec:algorithm-calc-terms}.
\item If the event was generated according to a tree-level matrix
  element, but had the highest multiplicity ($n=N$), then the event is
  reweighted by $ \prod_{i=1}^n \frac{\as(b\rho_i)}{\as(\mu)}
  \Delta_{S_{i-1}}(\rho_{i-1},\rho_i)$.
\end{itemize}
\item Continue the cascade below $\rho_{n+1}$
\end{enumerate}

\section{Results}
\label{sec:results}

Our algorithm has been implemented using \ariadne version 4.12
\cite{ARIADNE92}, which has been modified to include the possibility of
calculating the different weights needed. The matrix elements used is
taken from an implementation in \pythia version 6.414 \footnote{\pythia
has been modified to allow renormalization scales bigger than the center
of mass energy and to return negative weights instead of rounding to
zero for three partons at one loop.}
\cite{Sjostrand:2006za}, where the $e^+e^-$ matrix elements were
calculated in \cite{Ellis:1980wv} and parameterized in
\cite{Zhu:1983ps}.  The implementation includes the possibility of
generating $e^+e^-$ events according to zero, one, or two orders in
$\as$. This means that the four jets can only be generated according to
the tree-level matrix element, whereas the three jet contribution can be
generated using the one-loop contribution and the two jet can include up
to two loops. The different multiplicities are separated using a cutoff
in invariant mass divided by center of mass energy ($Q^2/s$), which can
be varied between 0.01 and 0.05. The process considered throughout this
section is $e^+e^-$ to hadrons at the $Z^0$ mass peak.

\FIGURE[t]{
    \epsfig{file=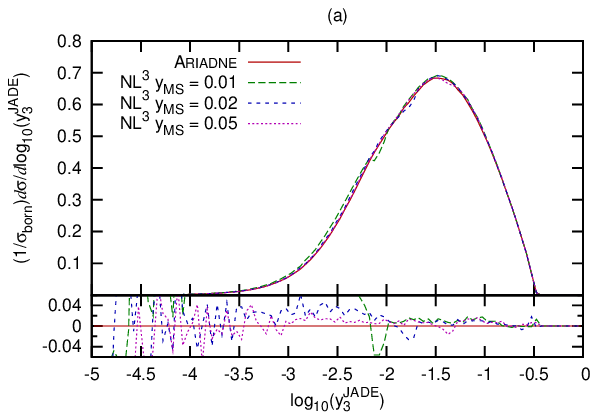,width=0.5\textwidth}%
    \epsfig{file=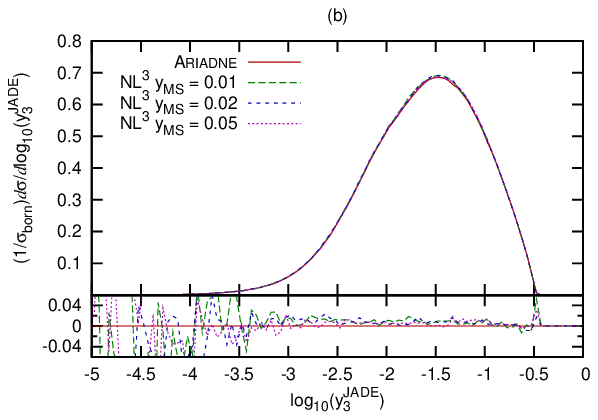,width=0.5\textwidth}\vspace*{-5mm}
    \caption{\label{fig:23corr} The parton-level $y_3$ spectra
      according to the JADE definition for the samples corrected with
      one-loop two-parton matrix element and tree-level three-parton
      matrix element.  (a) includes all flavours and masses and (b) is
      with only massless d-quarks. The in-sets at the bottom of the
      plots show the relative differences between the results from
      NL$^3$ and default \protect\ariadne, $(\sigma_{\mrm{NL^3}} -
      \sigma_{\mrm{ARIADNE}}) / \sigma_{\mrm{ARIADNE}}$.}
}

The first thing to be studied is how the algorithm behaves for the
somewhat trivial case of calculating three partons with a tree-level
matrix element and two partons to one loop. To study the effects of
the cutoff, the JADE\cite{Bartel:1986ua} jet clustering algorithm is
used, since it has a jet scale which closely resembles the scale used
for the cutoff, $\yms$. The matrix elements are calculated using a
fixed value of the $\as$ used in \ariadne at the renormalization
scale, which is set to $m_Z$ (\ie\ not using \eqref{eq:bparam} and
setting $b=1$ in the rest of section \ref{sec:theory} and
\ref{sec:algorithm}). The distribution in clustering scale for the
third jet for our new procedure\footnote{The results for our new
  procedure is throughout denoted NL$^3$.} is shown in figure
\ref{fig:23corr} with three different values of the merging scale,
0.01, 0.02 and 0.05 and is compared to the standard \ariadne
shower. The figure includes one plot with all flavours and masses and
one where only massless d-quarks was used. Including quark masses
there are some deviations, especially close to the cutoff. This is due
to slightly different treatments of the suppression of radiation from
heavy quarks (the dead-cone effect). The matrix element in \pythia uses
the exact formula, whereas in \ariadne a more general approximate
formula is used. However, when masses are not included no deviations
are visible, which should be the case since both the $K$-factor and
the matrix element correction are present in the shower to this order.

\FIGURE[t]{
    \epsfig{file=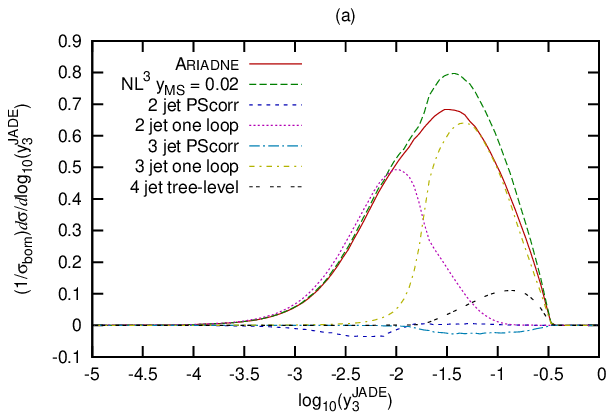,width=0.5\textwidth}%
    \epsfig{file=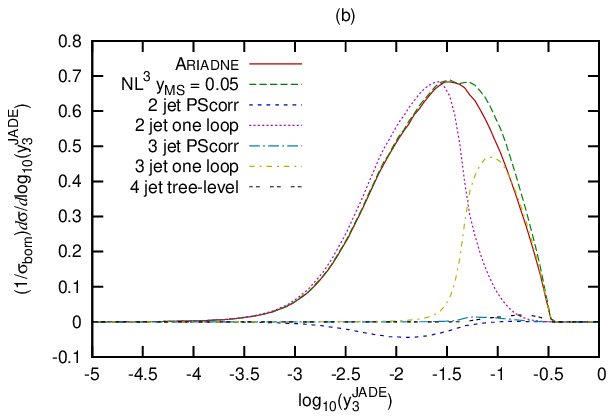,width=0.5\textwidth}\vspace*{-5mm}
    \caption{\label{fig:oneloop} The parton-level $y_3$ spectra according
to the JADE definition for the samples corrected with matrix elements
describing two partons at one loop, three partons at one loop and four
partons at tree level. (a) shows the curves for a cutoff of 0.02 and (b)
for a cutoff 0.05. Both figures include curves displaying the various
components of the NL$^3$ samples.}
}

Moving on to the simplest non-trivial case which includes two partons to
one loop, three partons to one loop and four partons to tree level. Also
here the clustering scale of the third jet using JADE is studied for the
same $\as$, renormalization scale and merging scales. The distribution
is shown in figure \ref{fig:oneloop} together with the various
components that make up each distribution. The cutoff 0.01 is not
included since it would have a negative two-parton cross section, which
is not allowed in the \pythia routines.

The different components in figure \ref{fig:oneloop} can be identified
in the following way. The curves marked \textit{one loop} are simply
the contributions from the one-loop matrix elements with a Sudakov
form factor according to equation (\ref{eq:1loopSud}) and a shower
added below the merging scale, the curves marked \textit{PScorr} are
the contribution calculated from the tree-level matrix elements
according to equation (\ref{eq:pscorr}) and the curve marked \textit{4
  jet tree-level} is the highest multiplicity contribution which
is calculated according to equation (\ref{eq:mepsinchmult}). The
dominant contributions are clearly the one-loop matrix elements for
two-parton and three-parton configurations. The four-parton matrix
element is also significant at the hard end of the spectrum, but the
contributions from the modified tree-level distributions are generally
small.

We also note that the modified tree-level contribution
(\textit{PScorr}) have a slightly negative value. This is a result of
the expansion of the Sudakov form factor together with the running
coupling.  However, negative weights can be avoided if one chooses a
merging scale defined using the ordering variable in the shower, and a
renormalization scale equal to the merging scale. As long as the
one-loop matrix elements are not themselves negative, which happens
for small enough merging scales, all events would then have positive
weights.

One important feature of the algorithm is the possibility of including
several multiplicities together. The importance of this is illustrated
in figure \ref{fig:oneloop} by the fact that the two-parton components
give a contribution which extend significantly above the merging scale
(while the opposite is true for the three parton
contribution). Clearly it would be problematic to try to describe the
jet distribution using only three- and four-parton matrix elements,
although the calculation would be formally correct to NLO
accuracy.

Another thing that is noticeable is that there is a significant overshoot
above the cutoff in figure \ref{fig:oneloop}. This is attributed to the
one-loop term of the three parton matrix element, which can not be
accurately reproduced in the cascade. The equivalent term in the cascade
is calculated using the Sudakov form factors and the value is
significantly smaller than the matrix element.

\FIGURE[t]{
    \epsfig{file=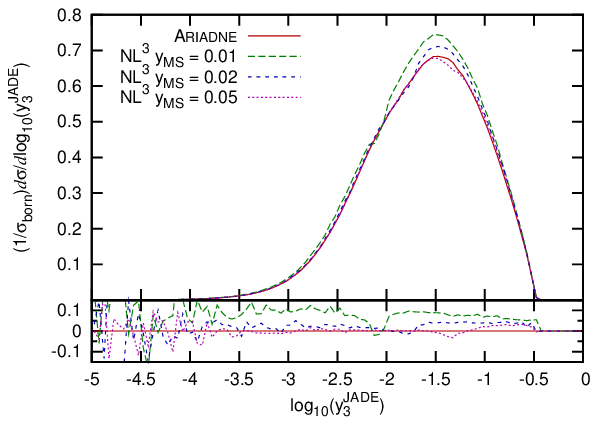,width=0.6\textwidth}\vspace*{-5mm}
    \caption{\label{fig:modalpha} The parton-level $y_3$ spectra
      according to the JADE definition for the NL$^3$ algorithm
      including a modified way of treating $\as$.}
}

The whole issue of the one-loop contribution being significantly larger
than the parton shower counterpart has another consequence, namely that
the $\as$ used in the shower is higher than what is fitted to precision
calculations, which include both fixed-order matrix elements and
logarithmic resummations. However, simply lowering $\as$ everywhere
would destroy the agreement between the curves below the cutoff. We
therefore both lower the value of $\as$ and modify the scale used as a
argument in the shower by using the parameter $b$ (defined in equation
(\ref{eq:bparam})). The value of $\as$ from the PDG
\cite{Amsler:2008zz} which is $\as(m_Z) = 0.1176$ which corresponds to
$\lqcd = 85.8$ MeV (assuming a leading order $\as$ which is used in
\ariadne), which should be compared to the \ariadne default $\lqcd =
220$ MeV.  Figure \ref{fig:modalpha} shows the curves using the PDG
value for $\as$ and $b= 85.8/220 = 0.389$. There is a much better
agreement for values above the cutoff, but there a still some
discrepancies.

\FIGURE[t]{
    \epsfig{file=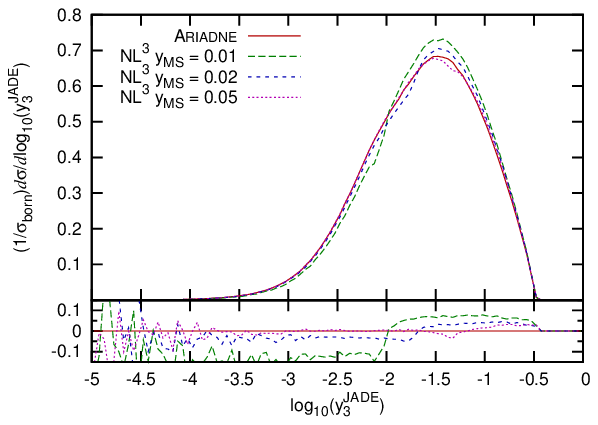,width=0.6\textwidth}\vspace*{-5mm}
    \caption{\label{fig:twoloop} The parton-level $y_3$ spectra according
to the JADE definition for the NL$^3$ algorithm including a modified way
of treating $\as$ and the two-parton matrix element at two loops.}
}
For the two-parton matrix element the calculation is also available at
two loops. This was discussed in general in section \ref{sec:twoloop}.
In figure \ref{fig:twoloop} the two-loop corrections have been included,
still using the same $\as$ treatment as described above. There is a clear
difference in that the curves no longer overlap with \ariadne for values
below the merging scale. This happens since the two-loop contribution is
beyond what can be reproduced by \ariadne.

\FIGURE[t]{
    \epsfig{file=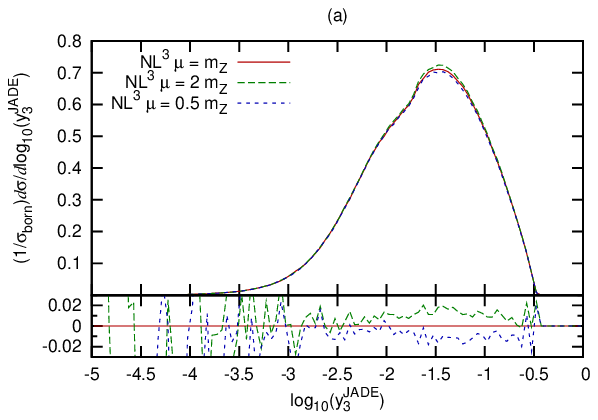,width=0.5\textwidth}%
    \epsfig{file=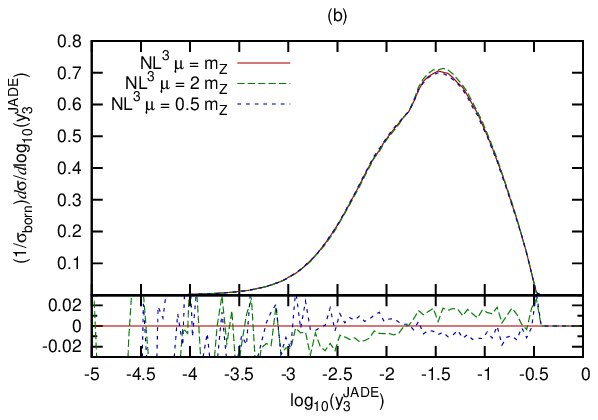,width=0.5\textwidth}\vspace*{-5mm}
    \caption{\label{fig:scale} The parton-level $y_3$ spectra according
to the JADE definition for the NL$^3$ algorithm including a modified way
of treating $\as$ where the renormalization scale has been varied up and
down by a factor of two. (a) includes the one-loop corrections and (b)
includes the two-parton matrix element at two loops.}
}

To check the consistency of the algorithm we have studied the
sensitivity to changes in the renormalization scale. In figure
\ref{fig:scale} the renormalization scale has been varied up and down
by a factor of two for both the case of one-loop correction and
including two partons at two loops. Overall, the sensitivity to
changes in the renormalization scale is small, with variations of
around two percent in the results. This is expected since all the
higher order terms comes from the shower which is unaffected by the
renormalization scale.

\FIGURE[t]{
    \epsfig{file=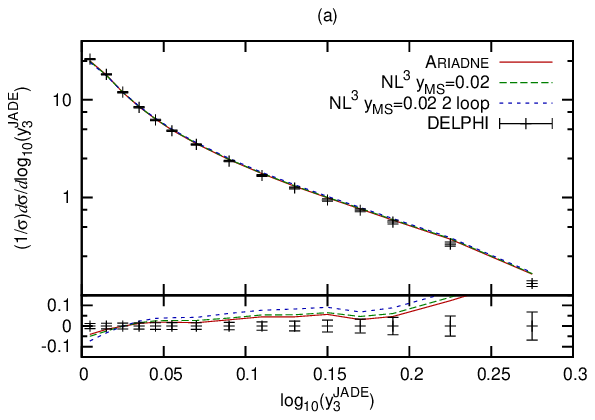,width=0.5\textwidth}%
    \epsfig{file=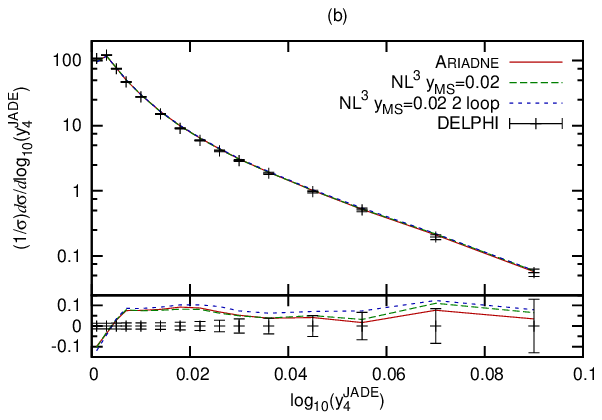,width=0.5\textwidth}
    \epsfig{file=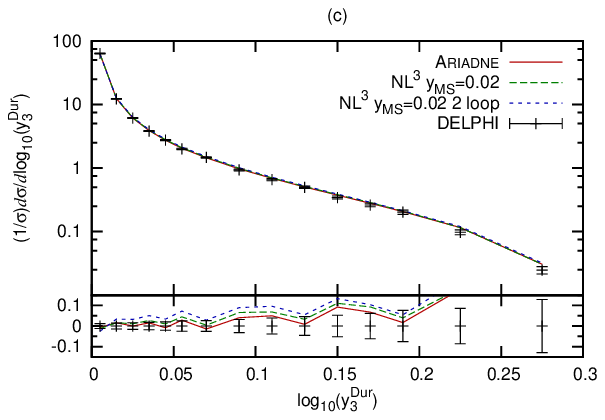,width=0.5\textwidth}%
    \epsfig{file=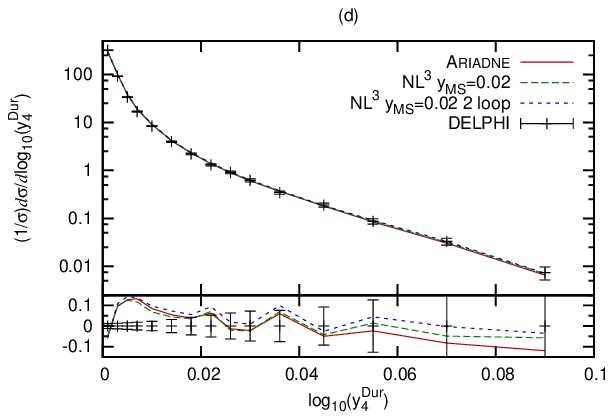,width=0.5\textwidth}\vspace*{-5mm}
    \caption{\label{fig:hadjet} Charged particle jet observables
      compared to DELPHI data and the standard \protect\ariadne shower
      for the NL$^3$ algorithm using a merging scale of 0.02.  The
      following jet observables are shown: (a) 3 jet JADE, (b) 4 jet
      JADE, (c) 3 jet Durham and (d) 4 jet Durham.}
  } \FIGURE[t]{
    \epsfig{file=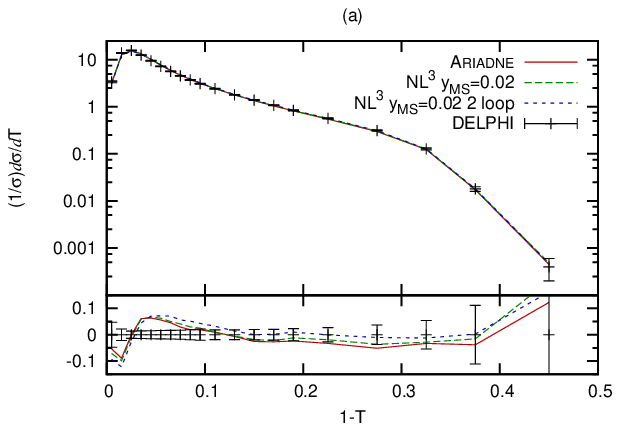,width=0.5\textwidth}%
    \epsfig{file=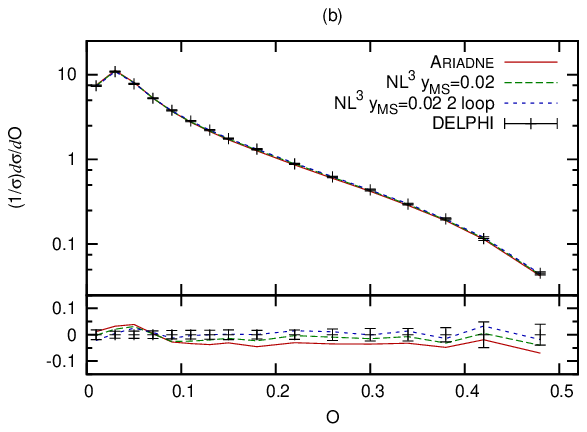,width=0.5\textwidth}\vspace*{-5mm}
    \caption{\label{fig:shape} The charged particle thrust and
      oblateness compared to DELPHI data and the standard \protect\ariadne
      shower for the NL$^3$ algorithm using a merging scale of 0.02.}
  }

Finally our algorithm is compared to data from the DELPHI
\cite{Abreu:1996na} experiment. It should be noted that all the data are
quite well reproduced by \ariadne and since we get small deviations in
the previous plot, we expect small differences here as well. The data to
be studied are all corrected to the particle level and includes only
charged particles. Figure \ref{fig:hadjet} shows the jet distributions
for the third and fourth jet according to the JADE \cite{Bartel:1986ua}
definition and the Durham \cite{Catani:1991hj} definition. The curves
include the central merging scale value of 0.02 with and without
the extra two-loop correction. All the results reproduce the data quite
well.

The results from the algorithm has also been compared to the shape
observables thrust and oblateness, which is shown in figure
\ref{fig:shape}. The agreement is again quite good. The curves
including the matrix element corrections actually do a bit better for
oblateness, which is to be expected since it is sensitive to
distributions including four jets where we have included the exact
tree-level matrix elements.

\section{Conclusions}
\label{sec:conclusions}

We have presented an algorithm for merging one-loop and tree-level
matrix elements with parton showers. The algorithm allows for the
inclusion of several different multiplicities, which is important for
simulating an entire process. For observables where the combination of
tree-level and one-loop matrix elements used gives the correct NLO
prediction, our procedure will also give correct NLO predictions but
with a resummation of leading logarithms which is of the same accuracy
as in the parton shower used. The basic principle of the procedure is
quite simple. The first two terms in the $\as$ expansion is subtracted
from the shower and the corresponding one-loop matrix element is
added. Although a simple idea, it leads to some complicated issues.

To calculate the first two terms in powers of $\as$ in a parton shower
requires that several different terms are taken into account. Both the
first term in the expansion of the Sudakov form factor and the first
term in the running coupling contributes at this level. In this paper
these terms were derived and subtracted from the shower. The calculation
was done within a framework similar to the CKKW-L method, but it is also
applicable to CKKW algorithms in general.

The modified shower is then added to a sample calculated
according to one-loop matrix elements. The requirement on the matrix
elements is that one should be able to specify a phase-space cut which
separates the parton multiplicities and decides whether or not a
parton is to be considered unresolved. This is not done within the
commonly used subtraction schemes, but can be solved by a modifying the
subtraction scheme outside the singular regions.

We have explicitly calculated all the weights without taking into
account the possibility of incoming hadrons, which is going to be the
topic of a future publication. We then implemented the procedure and
applied it to the process e$^+$e$^-$ to hadrons. The shower in
\ariadne was used and \pythia was used to generate the matrix
elements.

To test the consistency of the algorithm, jet distributions at parton
level were studied. The trivial case of two partons at one loop and
three partons at tree level was found to have only small deviations,
which disappeared if quark masses were excluded. The first nontrivial
case was how the algorithm behaves for two partons at one loop,
three partons at one loop and four partons at tree level. This led
to a clear overshoot due to the fact that the terms in the matrix
element is significantly bigger than those in the shower, which is
compensated by using a much larger $\as$ in the shower compared to
fits using matrix elements including loops. If the value of $\as$ is
adjusted according to our prescription, the agreement is quite good.

One important aspect of our procedure is the ability to combine
several different parton multiplicities in a consistent way. This
allows us to obtain corrections to a NLO prediction for a given
$n$-jet observable, stemming from cumulative sub-leading effects from
the parton shower added to the ($n-1$)-jet states.

Other aspects of the algorithms was explored, such as including the
two-parton matrix element to two-loop accuracy, which only led to
slight changes below the merging scale. The sensitivity to the choice
of renormalization scale was tested and a change of a factor of two in
the renormalization scale results in changes of around two percent in
the results.

The predictions of the algorithm has also been compared to four
different jets observables and two shape observables measured at LEP.
\ariadne already provides a good description of the data and including
the matrix element corrections gives similar agreement.

Overall the procedure has been shown to be consistent and give good
results at hadron level. These relatively simple cases establish a
proof of concept and a good staring point to explore the additional
pieces needed to simulate processes with incoming hadrons. If the
algorithm is developed further and the matrix element generators
improved, then there are good prospects for being able to merge
one-loop matrix elements and partons shower for the more interesting
LHC processes.

\section{Acknowledgments}

We thank Torbjörn Sjöstrand for useful discussions. Work supported in
part by the Marie Curie research training network ``MCnet'' (contract
number MRTN-CT-2006-035606).

\bibliographystyle{utcaps}
\bibliography{/home/shakespeare/people/leif/personal/lib/tex/bib/references,refs}

\end{document}

%% file: mydefs.tex
\usepackage{cite}
\usepackage{epsfig}
\usepackage{xspace}
\usepackage{graphics}
\usepackage[latin1]{inputenc}

\def\hide#1{}

\newcommand{\madevent}{M\scalebox{0.8}{AD}E\scalebox{0.8}{VENT}\xspace}

\newcommand{\herwig}{\scalebox{0.8}{HERWIG}\xspace}

\newcommand{\pythia}{P\scalebox{0.8}{YTHIA}\xspace}

\newcommand{\ariadne}{A\scalebox{0.8}{RIADNE}\xspace}

\newcommand{\as}{\ensuremath{\alpha_{\mathrm{s}}}}

\newcommand{\particle}[1]{\ensuremath{\mathrm{#1}}}

\newcommand{\el}{\particle{e}}

\newcommand{\ee}{\ensuremath{\el^+\el^-}}

\def\mrm#1{\mathrm{#1}}

\def\sup#1{\ensuremath{^{\mrm{#1}}}}

\def\sud#1{\ensuremath{\Delta_{S_{#1}}}}

\def\f2d3{\ensuremath{F_2^{\mrm{D}3}}}

\def\done#1{}

\providecommand{\eqref}[1]{eq.~(\ref{#1})\xspace}
\renewcommand{\eqref}[1]{eq.~(\ref{#1})\xspace}

\newcounter{aenumct}

\newcounter{ienumct}

\renewenvironment{itemize}{\begin{list}{$\bullet$}%
{\setlength{\topsep}{0mm}\setlength{\partopsep}{1mm}%
\setlength{\itemsep}{0mm}\setlength{\parsep}{1mm}}}%
{\end{list}}

\newcounter{enumct}
\renewenvironment{enumerate}{\begin{list}{\arabic{enumct}.}%
{\usecounter{enumct}\setlength{\topsep}{1mm}%
\setlength{\partopsep}{1mm}\setlength{\itemsep}{0mm}%
\setlength{\parsep}{1mm}}}{\end{list}}